\documentclass[twocolumn, prd, amssymb, preprintnumbers]{revtex4}

\def\CH{{\cal H}}

\def\IZ{\relax\ifmmode\mathchoice
{\hbox{\cmss Z\kern-.4em Z}}{\hbox{\cmss Z\kern-.4em Z}}
{\lower.9pt\hbox{\cmsss Z\kern-.4em Z}} {\lower1.2pt\hbox{\cmsss
Z\kern-.4em Z}}\else{\cmss Z\kern-.4em }\fi}
\def\IC{\relax\hbox{$\inbar\kern-.3em{\rm C}$}}
\def\IR{\relax{\rm I\kern-.18em R}}

\def\bT{{\bf T}}

\def\b{\beta}

\def\p{\pi}

\def\1{\relax 1 { \rm \kern-.35em I}}

\def\frac#1#2{{#1 \over #2}}

\def\p+{{\partial_+}}

\def\half{{1 \over 2}}
\def\ket#1{|#1\rangle}

\def\[{\left [}
\def\]{\right ]}
\def\({\left (}
\def\){\right )}

\def\CH{{\cal H}}

\def\IZ{\relax\ifmmode\mathchoice
{\hbox{\cmss Z\kern-.4em Z}}{\hbox{\cmss Z\kern-.4em Z}}
{\lower.9pt\hbox{\cmsss Z\kern-.4em Z}} {\lower1.2pt\hbox{\cmsss
Z\kern-.4em Z}}\else{\cmss Z\kern-.4em }\fi}
\def\IC{\relax\hbox{$\inbar\kern-.3em{\rm C}$}}
\def\IR{\relax{\rm I\kern-.18em R}}

\def\bT{{\bf T}}
\def\bS{{\bf S}}

\def\ajou#1&#2(#3){\ \sl#1\bf#2\rm(19#3)}

\def\TrH#1{ {\raise -.5em
                      \hbox{$\buildrel {\textstyle  {\rm Tr } }\over
{\scriptscriptstyle \CH _ {#1}}$}~}}

\def\CH{{\cal H}}

\def\IZ{\relax\ifmmode\mathchoice
{\hbox{\cmss Z\kern-.4em Z}}{\hbox{\cmss Z\kern-.4em Z}}
{\lower.9pt\hbox{\cmsss Z\kern-.4em Z}}
{\lower1.2pt\hbox{\cmsss Z\kern-.4em Z}}\else{\cmss Z\kern-.4em }\fi}
\def\IC{\relax\hbox{$\inbar\kern-.3em{\rm C}$}}
\def\IR{\relax{\rm I\kern-.18em R}}

\def\bT{{\bf T}}

\def\tS{{\tilde S}}
\def\tn{{\tilde n}}
\def\tw{{\tilde w}}

\def\b{\beta}

\def\p{\pi}

\def\1{\relax 1 { \rm \kern-.35em I}}

\def\frac#1#2{{#1 \over #2}}

\def\p+{{\partial_+}}

\def\half{{1 \over 2}}
\def\ket#1{|#1\rangle}

\def\[{\left [}
\def\]{\right ]}
\def\({\left (}
\def\){\right )}

\def\ajou#1&#2(#3){\ \sl#1\bf#2\rm(19#3)}

\def\TrH#1{ {\raise -.5em
                      \hbox{$\buildrel {\textstyle  {\rm Tr } }\over
{\scriptscriptstyle \CH _ {#1}}$}~}}

\newcommand{\bea}{\begin{eqnarray}}
\newcommand{\eea}{\end{eqnarray}}
\newcommand{\nn}{\nonumber}
\begin{document}
\preprint{TIFR/TH/05-44}
\preprint{CALT-68-2583}
\preprint{hep-th/0511120}
\title{Precision Microstate Counting of Small Black Rings}
\author{Atish Dabholkar$^{a}$, Norihiro Iizuka$^{a}$, Ashik
Iqubal$^a$,
 and Masaki Shigemori$^b$}
\affiliation{$^a$ Department of Theoretical Physics, Tata
Institute of Fundamental Research, Mumbai 400 005, India\\
$^b$California Institute of Technology 452-48, Pasadena, CA 91125,
USA}

\date{November 2005}

\begin{abstract}

We examine certain two-charge supersymmetric states with spin in
five-dimensional string theories which can be viewed as small
black rings when the gravitational coupling is large. Using the
4D-5D connection, these small black rings correspond to
four-dimensional non-spinning small black holes. Using this
correspondence, we compute the degeneracy of the microstates of
the small black rings exactly and show that it is in precise
agreement with the macroscopic degeneracy to all orders in an
asymptotic expansion.
Furthermore, we analyze the five-dimensional small black ring
geometry and show qualitatively that the Regge bound arises from
the requirement that closed time-like curves be absent.

\end{abstract}

\maketitle

\centerline{\bf {Introduction}}

One of the important successes of string theory has been that for
a special class of supersymmetric black holes with large classical
area in the supergravity approximation, one can explain the
Bekenstein-Hawking
entropy 
in terms of statistical counting of their microstates \cite{Strominger:1996sh}.
Recently, it has
become possible to extend these results to certain `small' black holes
that have vanishing classical entropy but the quantum corrected entropy
is nonzero and in precise agreement with the microscopic counting
\cite{Dabholkar:2004yr,Dabholkar:2004dq,Dabholkar:2005by,Dabholkar:2005dt,
Sen:1995in,Sen:2004dp,Sen:2005ch}.  In fact, for these small black
holes, a far more detailed comparison is possible and even the
subleading corrections to the entropy are found to be in agreement with
the state-counting to all orders in a perturbative expansion in large
charges. For this comparison to work, it is essential to include the
quantum corrections to the Bekenstein-Hawking area formula itself in a
systematic way \cite{Wald:1993nt, Iyer:1994ys} using the attractor
mechanism \cite{Ferrara:1995ih,
LopesCardoso:1998wt,LopesCardoso:1999cv,
LopesCardoso:1999xn,LopesCardoso:2000qm,Maldacena:1997de} and a specific
statistical ensemble \cite{Ooguri:2004zv}.%

The black hole entropy thus provides a valuable clue about the
microscopic structure of the theory. It is remarkable how tightly
constrained the structure of string theory is. Various terms in
the string effective action  have to be of a very definite form
with right coefficients in order that the resulting macroscopic
entropy matches with the counting of the microscopic quantum
states of the theory to all orders.

In this note we generalize these results to states that carry
spin. Spin introduces a number of qualitatively new features in
the analysis of spacetime geometry especially in conjunction with
supersymmetry. Supersymmetry requires that the angular velocity at
the horizon is zero because otherwise there would be an ergoregion
and energy can be extracted from the system for fixed charge in
conflict with BPS stability. Hence for supersymmetric states, the
angular momentum is typically swirling around outside the horizon.
The non-spinning horizons however can have nontrivial geometry or
topology. For example, adding spin to the D1-D5-P large black hole
in five dimensions \cite{Strominger:1996sh}, deforms the round
$\bf S^3$ horizon into an ellipsoid
\cite{Breckenridge:1996is,Gauntlett:1998fz}. The area of the
deformed horizon then correctly accounts for the modification of
the entropy due to spin. For the same system, in a different
regime of charge assignment, more exotic ring-like horizons with
$\bf S^2 \times S^1$ topology are possible
\cite{Elvang:2004rt,Elvang:2004ds,Gauntlett:2004wh,Gauntlett:2004qy,Bena:2004de}.
It is of interest therefore to know which of these possibilities
would be realized for states that correspond to small black holes
once spin is introduced and to ask if the counting still continues
to be in agreement with the macroscopic entropy.

We consider spinning Dabholkar-Harvey (DH) states
\cite{Dabholkar:1989jt,Dabholkar:1990yf} with two charges in
toroidally compactified heterotic string theory in five
dimensions. Using the chain of dualities and 4D-5D connection
\cite{Gaiotto:2005gf,Elvang:2005sa,Gaiotto:2005xt,Bena:2005ni}, it
was shown in \cite{Iizuka:2005uv}, that we can relate them to BPS
states in four dimensions with four charges but no spin. This
relation enables us to use the well-developed technology of the
attractor equations including certain quantum corrections in four
dimensions. One then finds that the resulting configuration has
finite entropy after including the quantum corrections
\cite{Iizuka:2005uv, Behrndt:2005he}. Moreover, as we will show,
the microscopic and macroscopic entropy in fact agree to all
orders in an asymptotic expansion.

\centerline{\bf{Small black rings and 4D-5D connection}}

Consider heterotic string compactified on $\bT^4 \times \bS^1$
where $\bT^4$ is a 4-torus in $\{6789\}$ directions and $\bS^1$ is
a circle along the $\{5\}$ direction. Consider now a string state
with winding number $w$ along the $X^5$ direction. In a given
winding sector, there is a tower of BPS states each in the
right-moving ground state but carrying arbitrary left-moving
oscillations subject to the Virasoro constraint $ N_L = 1 + nw$,
where $N_L$ is the left-moving oscillation number and  $n$ is the
quantized momentum along $X^5$
\cite{Dabholkar:1989jt,Dabholkar:1990yf}. Note that $N_L$ is
positive and hence a BPS state that satisfies this constraint has
positive $n$ for positive $w$ for large $N_L$. This state can
carry angular momentum $J$ say in the $\{34\}$ plane. The angular
momentum operator $J$ is given by
$
J = \sum_{n=1}^\infty \left[ a_{n}^\dagger a_{n} - {\bar
a}_{n}^\dagger {\bar a}_{n} \right],
$
where $a_n$ and ${\bar a}_n$ are the oscillator modes with frequency $n$
of the coordinates $(X^3 + i X^4)$ and $(X^3 - i X^4)$ respectively,
normalized as $[a_m,a_n^\dagger]=[\bar a_m,\bar
a_n^\dagger]=\delta_{mn}$.

Following the 4D-5D connection explained in \cite{Iizuka:2005uv}
we can map this state to a configuration in Type IIA compactified
on $\bf K3 \times T^2$ with charges D2-D2-D0-D4 which in turn is
dual to the DH states in heterotic string on $\bf T^4 \times S^1
\times \tS^1$ with momentum and winding $(n, w)$ and $(-\tn, \tw)$
along the circles $\bf S^1$ and $\bf \tS^1$ respectively, with all
integers $n, w, \tn, \tw$ positive. It is useful to state the
4D-5D connection entirely in the heterotic language. The basic
idea following
\cite{Gaiotto:2005gf,Elvang:2005sa,Gaiotto:2005xt,Bena:2005ni} is
to make use of the Taub-NUT geometry. For a Taub-NUT space with
unit charge, the geometry near the origin is $\bf R^4$ whereas at
asymptotic infinity it is $\bf R^3 \times \tS^1$. Thus a
contractible circle at the origin of Taub-NUT turns into a
non-contractible circle $\bf \tS^1$ at asymptotic infinity. The
angular momentum $J$ at the origin then turns into momentum
$\tilde n$ along the circle at infinity \cite{Elvang:2005sa}.

Consider now DH states with spin in heterotic string theory on
$\bf T^4 \times S^1$. Spinning strings that are wrapping along the
circle $\bf S^1$ in $\{5\}$ direction and rotating in the $\{34\}$
plane have a helical profile \cite{Dabholkar:1995nc,
Callan:1995hn, Lunin:2001fv,Lunin:2002qf}. The helix goes around a
contractible circle $\bf S^1_\psi$ of radius $R_\psi$ along an
angular coordinate $\psi$ in the $\{34\}$ plane as the string
wraps around the non-contractible circle $\bf S^1$. Let us denote
the pitch of the helix by $p$, which is the winding number of the
projection of the  helix onto the contractible circle.
Macroscopically it corresponds to a dipole charge.
We can now embed this system in Taub-NUT space with very large
Taub-NUT radius $R_{\rm TN} \gg R_\psi$ and regard $\bf S^1_\psi$
in $\bf R^4$ as being situated at the origin of a Taub-NUT
geometry. Varying the radius of Taub-NUT, which is a modulus, we
can smoothly go to the regime $R_{\rm TN} \ll R_\psi$. Then the
contractible $\bf S^1_\psi$ effectively turns into the
non-contractible circle $\bf \tS^1$ at asymptotic infinity. The
entropy of BPS states is not expected to change under such an
adiabatic change of moduli. We can dimensionally reduce the system
to 4D along $\bf \tS^1$ and obtain a 4D DH state with four charges
$(n, w, -\tn, \tw)$ with the identification that $\tn =J$ and $\tw
=p$.
This system has a string scale horizon in 4D
\cite{Dabholkar:2004yr,Dabholkar:2004dq,Dabholkar:2005by,Dabholkar:2005dt,
Sen:1995in,Sen:2004dp,Sen:2005ch} which  suggests that the
original spinning DH system in 5D is a small black ring
\cite{Iizuka:2005uv}.

Since we have unit Taub-NUT charge to begin with, we do not have a
purely electric configuration in 4D but instead have a Kaluza-Klein
monopole of unit charge in addition to the 4-charge purely electric
small black hole.  However, since the helix is far away from the origin
of Taub-NUT space in 5D before dimensional reduction, the KK-monopole is
sitting far away from the 4-charge black hole in 4D\@. The separation is
determined by Denef's constraint \cite{Denef:2000nb} and is determined
by $J$ and the asymptotic values of the moduli and can thus be made
arbitrarily large. The local microscopic counting therefore does not
depend on the addition of the KK-monopole and is given by the counting
of DH states.

\centerline{\bf{Four-dimensional counting}}

In the 4D description, the state is specified by the charge vector $Q$
in the Narain charge lattice $\Gamma^{2,2}$ of the $\bf S^1
\times \tS^1$ factor with four integer entries. The norm of this vector
is
\begin{eqnarray}\label{charge}
{Q^2 \over 2} &=& \half \left(%
\begin{array}{cccc}
  n & w &- \tn &\tw \\
\end{array}%
\right)\left(%
\begin{array}{cccc}
  0 & 1& 0 & 0 \\
  1 & 0 & 0 & 0 \\
 0 & 0 & 0& 1 \\
  0& 0& 1 & 0 \\
\end{array}%
\right)
\left(%
\begin{array}{c}
  n \\
  w \\
  -\tn\\
  \tw \\
\end{array}%
\right) \,
\end{eqnarray}
The degeneracy of these perturbative DH states can be computed
exactly and the asymptotic degeneracy for large $Q^2$ is given as
in \cite{Dabholkar:2005by,Dabholkar:2005dt} by
\begin{equation}\label{micro4D}
\Omega_{\rm micro} (n, w, \tn, \tw )
 \sim {\hat I}_{13}(4\pi \sqrt{Q^2\over 2})
 = {\hat I}_{13}(4\pi \sqrt{n w- \tn \tw}) \,,
\end{equation}
where ${\hat I}_{13}(z)$ is the modified Bessel function defined
in \cite{Dabholkar:2005dt}.

Turning to the macroscopic degeneracy, we compute it using the OSV
relation \cite{Ooguri:2004zv} between topological string partition
function and the macroscopic degeneracy, for the non-spinning
four-dimensional configuration. For this purpose we use the Type
IIA description, viewing this state as a collection of D2-D2-D0-D4
branes. There are $n$ D2-branes wrapping a 2-cycle $\alpha_1$ in
$\bf K3$ and $w$ D2-branes wrapping a 2-cycle $\alpha_2$ such that
the intersection matrix of $\alpha_1$ and $\alpha_2$ is as in the
upper-left $2\times 2$ block of the matrix in (\ref{charge}). We
therefore identify the charges as $(q_2, q_3) = (n, w)$.
Similarly, we identify the D0-D4 charges as $(q_0, p^1)= (\tn,
\tw)$ so in the labelling of charges used in
\cite{Dabholkar:2005dt}, we have $Q=(q_2, q_3, q_0, p^1)$ and all
other charges zero. Using the formula (2.26) in
\cite{Dabholkar:2005dt} we then see that the macroscopic
degeneracy is given by
\begin{eqnarray}\label{macro4D}
\Omega_{\rm macro} (n, w, J ) &\sim& (p^1)^2 {\hat I}_{13}(4\pi
\sqrt{q_2\,q_3 -p^1 q_0}) \,\nn\\
&=&  (p^1)^2 {\hat I}_{13}(4\pi \sqrt{n w- \tn \tw}) \,.
\end{eqnarray}
Therefore, up to the overall $(p^1)^2$ factor, the microscopic
(\ref{micro4D}) and macroscopic (\ref{macro4D})
degeneracies match
precisely to all orders in an asymptotic expansion for large
$Q^2$.

\centerline{\bf{Five-dimensional microscopic counting}}

We now would like to count the degeneracy of the spinning DH
system from the 5D side.  The nontrivial issue is to determine the
correct ensemble.
The relevant states correspond to quantum fluctuations around a specific
coherent oscillating state which is essentially Bose-Einstein condensate
on the worldsheet and describes the helical geometry with pitch $p$
\cite{Iizuka:2005uv}.
As we will argue below, the precise microstates turn out to be of the
form
\bea \label{microstates}
\underbrace{(a_{p}^\dagger)^J \raisebox{-3.25ex}{\rule{0pt}{1ex}} }_
{\begin{minipage}{16ex}\scriptsize
   microscopic origin of $\tn=J$ and $\tw =p$ of the ring
  \end{minipage}}
  \!\! \times\,\,\,\,
 \underbrace{\prod_{n=1}^\infty \left[\prod_{i=1,2,\pm,5...24}
 (\alpha_{-n}^{i})^{N_{ni}}\right]\ket{0}}_
 {\begin{minipage}{26ex}\scriptsize
   fluctuation: all possible states with level $N_{\rm eff} \equiv
   N-pJ$.  Angular momentum $J$ is not fixed.
  \end{minipage}}
\eea
for $J,p>0$.  Namely, {\em we consider the states with the worldsheet
energy $N_{\rm eff} \equiv N-pJ$ and chemical potential $\mu$ conjugate
to $J$ set to zero, and multiply all those states by
$(a_{p}^\dagger)^J$.}  Note in particular that the $a_p^\dagger$ and
$\bar a_p^{\dagger}$ oscillators are included in the fluctuation part.
The degeneracy of the states (\ref{microstates}) is the same as that of
the DH system with $Q^2/2=N_{\rm eff}$ and proportional to
$\hat{I}_{13}(4 \pi \sqrt{N-p J})$, in precise agreement with
(\ref{micro4D}), (\ref{macro4D}).
If $J,p<0$, $(a_{p}^\dagger)^J$ in
(\ref{microstates}) must be replaced by $(\bar a_{|p|}^\dagger)^{|J|}$.

This separation between the classical coherent condensate that
describes the large helix and the small quantum fluctuations
around it that account for the entropy is similar to the one used
in
\cite{Palmer:2004gu,Marolf:2004fy,Bak:2004kz,Bena:2004tk,Kraus:2005vz}.
It is valid in the regime when $R_\psi$ is much larger than the
amplitude of fluctuation.
Therefore, we conclude that in
this regime, the states of the form (\ref{microstates})  are the
states that account for the microstates of the ring. This in turn
agrees with (\ref{micro4D}) and (\ref{macro4D}) in 4D through the
4D-5D connection.

Note that (\ref{microstates}) means that the microscopic counting
in 5D must {\em not\/} be done for fixed angular momentum $J$.
Fixing $J$ would impose an additional constraint on the
fluctuation part in (\ref{microstates}). From the 4D point of
view,  it would correspond to imposing a constraint on the
worldsheet oscillators of the DH system, which would lead to a
result contradictory to the 4D degeneracy (\ref{micro4D}),
(\ref{macro4D}).
To demonstrate this, let us count the degeneracy of spinning DH
states with {\em fixed $J$}.  The degeneracy $\Omega(N,J)$ is
summarized in the partition function
\bea
\label{partition}
Z(\beta, \mu) = \sum_{N,J} \Omega(N, J) \, q^N c^J,
\quad q = e^{-\beta}, \quad c = e^{\beta \mu},
\eea
where $N \equiv n \,w = N_L -1$.  $\beta$ can be thought as the inverse
temperature on the worldsheet for a $1+1$ gas of left-moving $24$ bosons
conjugate to the total energy $N$ and $\mu$ can be thought of as the
chemical potential conjugate to the quantum number $J$ of this gas.
Since $N_L$ is the oscillation number for the $24$ left-moving
transverse bosons, using the expression $J = \sum_{n=1}^\infty
\left[ a_{n}^\dagger a_{n} - {\bar a}_{n}^\dagger {\bar a}_{n}
\right]$, the partition function can be readily evaluated
\cite{Russo:1994ev} and is given by
\bea
\label{partitionfinal}
Z(\beta, \mu)
&=&
\biggl[ q \prod_{n=1}^{\infty} (1 - q^n)^{22}
(1 - c \, q^n) (1 - c^{-1}q^n) \biggr] ^{-1}
\nn\\
&=& {1 \over
{\eta^{21}(e^{-\beta}) 
}}
{{2 i
 \sinh(\beta \mu/2) } \over
{\theta_{11}(\beta\mu/ 2 \pi i,i\beta/2 \pi)}} \,,
\eea
in terms of the standard Dedekind eta function and theta function
with characteristics.

The number of states with given $N$ and $J$ is then given by the inverse
Laplace transform:
%
$    \Omega(N, J) =
    {1\over (2\pi i)^2}
    \int_{C_\beta} \!\! d\b\, e^{\b N}\!
    \int_{C_\mu} \!\! \beta\,d\mu\, e^{-\mu \beta J}
    Z(\beta, \mu),~
$
%
where the contour $C_\beta$ runs from $-i \pi + \gamma$ to $+i \pi
+ \gamma$ with $\gamma > 0$ to avoid singularities on the
imaginary axis. Similarly, $C_\mu$ goes from $-\pi
i/\beta+\epsilon$ to $+\pi i/\beta+\epsilon$ with $-1<\epsilon<1$
to avoid poles.  To find the  asymptotic degeneracy at large $N$,
we want to take the high temperature limit, or $\beta \rightarrow
0$.  Using the  modular properties of the Dedekind eta and the
theta functions we can write the degeneracy at high temperature as
in \cite{Russo:1994ev} as
\bea
\label{final}
\Omega(N, J) &\sim& {1\over 2\pi i} \int_{C_\beta} \!\!
d\beta \, e^{\beta N + {(2\pi)^2/\beta}}
\left({\beta \over 2\pi}\right)^{\!12}\!
I(\beta, J)  \,,~~
\eea
where $I(\beta, J)$ is defined by 
\bea
I(\beta,J)
&=& {1 \over 2 \pi i}
\int_{C_\mu} \!\!
d\mu \, e^{ - \beta \mu^2/2- \beta \mu J} \,
{{\sinh(\beta \mu/2) } \over \sin (\pi\mu)} \,.  ~
 \label{nuintegral}
\eea
To arrive at (\ref{final}), we dropped terms that are
exponentially suppressed for small $\beta$ as $e^{-(2
\pi)^2/\beta}$.  This is justified  although $\beta$ is still to
be integrated over, because the saddle point around
$\beta\sim1/\sqrt{N-J}\ll 1$ will make the leading contribution,
as we will see below.

Now we evaluate (\ref{nuintegral}) using the method of residues. Deform
the contour $C_\mu$ into sum of three intervals $C_1=[-\pi
i/\beta+\epsilon,-\pi i/\beta+K]$, $C_2=[-\pi i/\beta+K,\pi i/\beta+K]$,
and $C_3=[\pi i/\beta+K,\pi i/\beta+\epsilon]$, with $K\gg 1$.  One can
readily show that the contour integral along $C_{1,2,3}$
vanishes due to the periodicity of the original integrand (\ref{partitionfinal})
in the small $\beta$ and large $K$ limit.  In the process of
deforming the contour, we pick up poles at $\mu = m$, $m= 1, 2,
\ldots$. In the end, we obtain
\bea\label{pole}
 I(\beta,J) &\sim&
-\mathop{\rm Res}\limits_{\mu=1}
\biggl[e^{ - \beta \mu^2/2- \beta \mu J}\,
{{\sinh (\beta \mu/2) } \over \sin (\pi \mu)} \biggr]
+
{\cal{O}}(e^{- 2 \beta J}) \nn \\
&\sim&   (1 - e^{ - \beta})
 \, e^{- \beta J} + {\cal{O}}(e^{- 2 \beta J}) \,.
\eea
Here ${\cal{O}}(e^{- 2 \beta J})$ comes from the poles at $\mu =
2,3,\cdots$ and negligible when $J = {\cal{O}} (N)$ since $\beta J =
{\cal{O}} (N^{1/2})$.
We interpret the term $\propto e^{- \beta J}$ as the contribution
from the $p=1$ sector.
Substituting this back into (\ref{final}), we
conclude that the degeneracy $ \Omega(N, J)$ is
$\sim{1\over 2\pi i} \int \!d\beta ({\beta / 2\pi})^{\!12}\!
(\beta - {\beta^2 \over 2}  + 
+\cdots ) \,
e^{{(2\pi)^2/ \beta} + {\beta(N-J)}} \,.
$
%
Each term in the integral is of the Bessel type as discussed in
\cite{Dabholkar:2005dt}, and thus the final result is
$\Omega (N, J) \sim {\hat I}_{14} (4\pi \sqrt{N-J}) - {(2\pi / 2)} {\hat
I}_{15} (4\pi \sqrt{N-J}) + \cdots
$, which agrees with (\ref{micro4D}) and
(\ref{macro4D}) with $p=1$ only in the leading exponential but disagrees
in the subleading corrections.  This demonstrates that microscopic
counting in 5D must be done not for fixed $J$ but for the states
(\ref{microstates}).
One can show that states with fixed $\mu\neq0$ also lead to degeneracy
in disagreement with (\ref{micro4D}), (\ref{macro4D}).

In general, subleading corrections to thermodynamic quantities depend on
the choice of the statistical ensembles and are different for different
ensembles. For example, even for non-spinning black holes the ensemble
with fixed angular momentum $J=0$ differs from the ensemble with fixed
chemical potential $\mu=0$ in subleading corrections. 
It was noted in \cite{Dabholkar:2005dt} that the correct microscopic ensemble that is
consistent with the OSV conjecture is the one with $\mu=0$.  In our
case, the description of small black rings requires that we are also
fixing the pitch of the helix as an additional requirement and that we
are counting states around this classical coherent condensate on the
worldsheet.

\centerline{\bf{Five-dimensional macroscopic geometry}}
Finally we comment on the geometry of this 5D small black ring, which
can be determined in the near ring limit 
by using the 4D-5D uplift.  Exact uplift is possible near the
horizon, since the near horizon geometry of 4D black hole can be
determined precisely by using string-corrected attractor equations. 
Let us consider the case where $p \, J = {\cal{O}}(N)$. We consider a
helix with general pitch $p$ even though the contribution from the $p =
1$ subsector dominates entropy.  In 4D heterotic string theory, the near-horizon black hole
geometry is determined by the attractor equations to be
$e^{2\phi_4} \sim 1/\sqrt{N-pJ} \,$,
$\sqrt{g_{\psi\psi}} \sim \sqrt{J/ p}$
%
and the horizon radius is $r_{\bf S^2} \sim l_{s}$.  Using $e^{2 \phi_5}
\sim e^{2 \phi_4} \sqrt{g_{\psi\psi}}$ and
$l_{s} \sim e^{-2\phi_5/3} l_{pl}^{(5)}$,
the scale of $\bf S^1_\psi$ along ring is given by
\bea
r_{\bf S^1} &\sim & \sqrt{g_{\psi\psi}} \, l_s
\sim p^{-1/3}J^{1/3}(N - p J)^{1/6} \, l^{(5)}_{pl}
\,, \, \nn \\
r_{\bf S^2} &\sim& l_s \sim  p^{1/6} J^{-1/6}(N-p J)^{1/6} l_{pl}^{(5)} \,, 
\eea
such that $S \sim A/4G_{5} \sim r_{\bf S^1} (r_{\bf
S^2})^2/(l_{pl}^{(5)})^3 \sim \sqrt{N- p J}$.  One important qualitative
feature of the solution is that when $p J$ exceeds $N$, the solution
develops closed timelike curves.
When $N \ge p J$ is saturated, $g_{\psi\psi} \ge 0$ is saturated
at the ring horizon.
Hence, the Regge bound $N \ge p J$ on
the angular momentum of the underlying microstates can be understood
from the macroscopic solution as a consequence of the physical
requirement that closed timelike curves be absent.  The details will be presented elsewhere \cite{Progress}.
%
\centerline{\bf Acknowledgments}
We are very grateful to Ashoke Sen for numerous illuminating discussions
and collaboration on \cite{Progress}.
A.D. would like to thank the high energy
group at ASICTP, Trieste for their hospitality where part of this work
was completed.
The work of M.S. was supported in part by Department of
Energy grant DE-FG03-92ER40701 and a Sherman Fairchild Foundation
postdoctoral fellowship.

\bibliography{ringref}

\begin{thebibliography}{44}
\expandafter\ifx\csname natexlab\endcsname\relax\def\natexlab#1{#1}\fi
\expandafter\ifx\csname bibnamefont\endcsname\relax
  \def\bibnamefont#1{#1}\fi
\expandafter\ifx\csname bibfnamefont\endcsname\relax
  \def\bibfnamefont#1{#1}\fi
\expandafter\ifx\csname citenamefont\endcsname\relax
  \def\citenamefont#1{#1}\fi
\expandafter\ifx\csname url\endcsname\relax
  \def\url#1{\texttt{#1}}\fi
\expandafter\ifx\csname urlprefix\endcsname\relax\def\urlprefix{URL }\fi
\providecommand{\bibinfo}[2]{#2}
\providecommand{\eprint}[2][]{\url{#2}}

\bibitem[{\citenamefont{Strominger and Vafa}(1996)}]{Strominger:1996sh}
\bibinfo{author}{\bibfnamefont{A.}~\bibnamefont{Strominger}} \bibnamefont{and}
  \bibinfo{author}{\bibfnamefont{C.}~\bibnamefont{Vafa}},
  \bibinfo{journal}{Phys. Lett.} \textbf{\bibinfo{volume}{B379}},
  \bibinfo{pages}{99} (\bibinfo{year}{1996}), \eprint{hep-th/9601029}.

\bibitem[{\citenamefont{Dabholkar}(2005)}]{Dabholkar:2004yr}
\bibinfo{author}{\bibfnamefont{A.}~\bibnamefont{Dabholkar}},
  \bibinfo{journal}{Phys. Rev. Lett.} \textbf{\bibinfo{volume}{94}},
  \bibinfo{pages}{241301} (\bibinfo{year}{2005}), \eprint{hep-th/0409148}.

\bibitem[{\citenamefont{Dabholkar et~al.}(2004)\citenamefont{Dabholkar,
  Kallosh, and Maloney}}]{Dabholkar:2004dq}
\bibinfo{author}{\bibfnamefont{A.}~\bibnamefont{Dabholkar}},
  \bibinfo{author}{\bibfnamefont{R.}~\bibnamefont{Kallosh}}, \bibnamefont{and}
  \bibinfo{author}{\bibfnamefont{A.}~\bibnamefont{Maloney}},
  \bibinfo{journal}{JHEP} \textbf{\bibinfo{volume}{12}}, \bibinfo{pages}{059}
  (\bibinfo{year}{2004}), \eprint{hep-th/0410076}.

\bibitem[{\citenamefont{Dabholkar
  et~al.}(2005{\natexlab{a}})\citenamefont{Dabholkar, Denef, Moore, and
  Pioline}}]{Dabholkar:2005by}
\bibinfo{author}{\bibfnamefont{A.}~\bibnamefont{Dabholkar}},
  \bibinfo{author}{\bibfnamefont{F.}~\bibnamefont{Denef}},
  \bibinfo{author}{\bibfnamefont{G.~W.} \bibnamefont{Moore}}, \bibnamefont{and}
  \bibinfo{author}{\bibfnamefont{B.}~\bibnamefont{Pioline}},
  \bibinfo{journal}{JHEP} \textbf{\bibinfo{volume}{08}}, \bibinfo{pages}{021}
  (\bibinfo{year}{2005}{\natexlab{a}}), \eprint{hep-th/0502157}.

\bibitem[{\citenamefont{Dabholkar
  et~al.}(2005{\natexlab{b}})\citenamefont{Dabholkar, Denef, Moore, and
  Pioline}}]{Dabholkar:2005dt}
\bibinfo{author}{\bibfnamefont{A.}~\bibnamefont{Dabholkar}},
  \bibinfo{author}{\bibfnamefont{F.}~\bibnamefont{Denef}},
  \bibinfo{author}{\bibfnamefont{G.~W.} \bibnamefont{Moore}}, \bibnamefont{and}
  \bibinfo{author}{\bibfnamefont{B.}~\bibnamefont{Pioline}}
  (\bibinfo{year}{2005}{\natexlab{b}}), \eprint{hep-th/0507014}.

\bibitem[{\citenamefont{Sen}(1995)}]{Sen:1995in}
\bibinfo{author}{\bibfnamefont{A.}~\bibnamefont{Sen}}, \bibinfo{journal}{Mod.
  Phys. Lett.} \textbf{\bibinfo{volume}{A10}}, \bibinfo{pages}{2081}
  (\bibinfo{year}{1995}), \eprint{hep-th/9504147}.

\bibitem[{\citenamefont{Sen}(2005{\natexlab{a}})}]{Sen:2004dp}
\bibinfo{author}{\bibfnamefont{A.}~\bibnamefont{Sen}}, \bibinfo{journal}{JHEP}
  \textbf{\bibinfo{volume}{05}}, \bibinfo{pages}{059}
  (\bibinfo{year}{2005}{\natexlab{a}}), \eprint{hep-th/0411255}.

\bibitem[{\citenamefont{Sen}(2005{\natexlab{b}})}]{Sen:2005ch}
\bibinfo{author}{\bibfnamefont{A.}~\bibnamefont{Sen}}
  (\bibinfo{year}{2005}{\natexlab{b}}), \eprint{hep-th/0504005}.

\bibitem[{\citenamefont{Wald}(1993)}]{Wald:1993nt}
\bibinfo{author}{\bibfnamefont{R.~M.} \bibnamefont{Wald}},
  \bibinfo{journal}{Phys. Rev.} \textbf{\bibinfo{volume}{D48}},
  \bibinfo{pages}{3427} (\bibinfo{year}{1993}), \eprint{gr-qc/9307038}.

\bibitem[{\citenamefont{Iyer and Wald}(1994)}]{Iyer:1994ys}
\bibinfo{author}{\bibfnamefont{V.}~\bibnamefont{Iyer}} \bibnamefont{and}
  \bibinfo{author}{\bibfnamefont{R.~M.} \bibnamefont{Wald}},
  \bibinfo{journal}{Phys. Rev.} \textbf{\bibinfo{volume}{D50}},
  \bibinfo{pages}{846} (\bibinfo{year}{1994}), \eprint{gr-qc/9403028}.

\bibitem[{\citenamefont{Ferrara et~al.}(1995)\citenamefont{Ferrara, Kallosh,
  and Strominger}}]{Ferrara:1995ih}
\bibinfo{author}{\bibfnamefont{S.}~\bibnamefont{Ferrara}},
  \bibinfo{author}{\bibfnamefont{R.}~\bibnamefont{Kallosh}}, \bibnamefont{and}
  \bibinfo{author}{\bibfnamefont{A.}~\bibnamefont{Strominger}},
  \bibinfo{journal}{Phys. Rev.} \textbf{\bibinfo{volume}{D52}},
  \bibinfo{pages}{5412} (\bibinfo{year}{1995}), \eprint{hep-th/9508072}.

\bibitem[{\citenamefont{Lopes~Cardoso et~al.}(1999)\citenamefont{Lopes~Cardoso,
  de~Wit, and Mohaupt}}]{LopesCardoso:1998wt}
\bibinfo{author}{\bibfnamefont{G.}~\bibnamefont{Lopes~Cardoso}},
  \bibinfo{author}{\bibfnamefont{B.}~\bibnamefont{de~Wit}}, \bibnamefont{and}
  \bibinfo{author}{\bibfnamefont{T.}~\bibnamefont{Mohaupt}},
  \bibinfo{journal}{Phys. Lett.} \textbf{\bibinfo{volume}{B451}},
  \bibinfo{pages}{309} (\bibinfo{year}{1999}), \eprint{hep-th/9812082}.

\bibitem[{\citenamefont{Lopes~Cardoso
  et~al.}(2000{\natexlab{a}})\citenamefont{Lopes~Cardoso, de~Wit, and
  Mohaupt}}]{LopesCardoso:1999cv}
\bibinfo{author}{\bibfnamefont{G.}~\bibnamefont{Lopes~Cardoso}},
  \bibinfo{author}{\bibfnamefont{B.}~\bibnamefont{de~Wit}}, \bibnamefont{and}
  \bibinfo{author}{\bibfnamefont{T.}~\bibnamefont{Mohaupt}},
  \bibinfo{journal}{Fortsch. Phys.} \textbf{\bibinfo{volume}{48}},
  \bibinfo{pages}{49} (\bibinfo{year}{2000}{\natexlab{a}}),
  \eprint{hep-th/9904005}.

\bibitem[{\citenamefont{Lopes~Cardoso
  et~al.}(2000{\natexlab{b}})\citenamefont{Lopes~Cardoso, de~Wit, and
  Mohaupt}}]{LopesCardoso:1999xn}
\bibinfo{author}{\bibfnamefont{G.}~\bibnamefont{Lopes~Cardoso}},
  \bibinfo{author}{\bibfnamefont{B.}~\bibnamefont{de~Wit}}, \bibnamefont{and}
  \bibinfo{author}{\bibfnamefont{T.}~\bibnamefont{Mohaupt}},
  \bibinfo{journal}{Class. Quant. Grav.} \textbf{\bibinfo{volume}{17}},
  \bibinfo{pages}{1007} (\bibinfo{year}{2000}{\natexlab{b}}),
  \eprint{hep-th/9910179}.

\bibitem[{\citenamefont{Lopes~Cardoso
  et~al.}(2000{\natexlab{c}})\citenamefont{Lopes~Cardoso, de~Wit, Kappeli, and
  Mohaupt}}]{LopesCardoso:2000qm}
\bibinfo{author}{\bibfnamefont{G.}~\bibnamefont{Lopes~Cardoso}},
  \bibinfo{author}{\bibfnamefont{B.}~\bibnamefont{de~Wit}},
  \bibinfo{author}{\bibfnamefont{J.}~\bibnamefont{Kappeli}}, \bibnamefont{and}
  \bibinfo{author}{\bibfnamefont{T.}~\bibnamefont{Mohaupt}},
  \bibinfo{journal}{JHEP} \textbf{\bibinfo{volume}{12}}, \bibinfo{pages}{019}
  (\bibinfo{year}{2000}{\natexlab{c}}), \eprint{hep-th/0009234}.

\bibitem[{\citenamefont{Maldacena et~al.}(1997)\citenamefont{Maldacena,
  Strominger, and Witten}}]{Maldacena:1997de}
\bibinfo{author}{\bibfnamefont{J.~M.} \bibnamefont{Maldacena}},
  \bibinfo{author}{\bibfnamefont{A.}~\bibnamefont{Strominger}},
  \bibnamefont{and} \bibinfo{author}{\bibfnamefont{E.}~\bibnamefont{Witten}},
  \bibinfo{journal}{JHEP} \textbf{\bibinfo{volume}{12}}, \bibinfo{pages}{002}
  (\bibinfo{year}{1997}), \eprint{hep-th/9711053}.

\bibitem[{\citenamefont{Ooguri et~al.}(2004)\citenamefont{Ooguri, Strominger,
  and Vafa}}]{Ooguri:2004zv}
\bibinfo{author}{\bibfnamefont{H.}~\bibnamefont{Ooguri}},
  \bibinfo{author}{\bibfnamefont{A.}~\bibnamefont{Strominger}},
  \bibnamefont{and} \bibinfo{author}{\bibfnamefont{C.}~\bibnamefont{Vafa}}
  (\bibinfo{year}{2004}), \eprint{hep-th/0405146}.

\bibitem[{\citenamefont{Breckenridge et~al.}(1997)\citenamefont{Breckenridge,
  Myers, Peet, and Vafa}}]{Breckenridge:1996is}
\bibinfo{author}{\bibfnamefont{J.~C.} \bibnamefont{Breckenridge}},
  \bibinfo{author}{\bibfnamefont{R.~C.} \bibnamefont{Myers}},
  \bibinfo{author}{\bibfnamefont{A.~W.} \bibnamefont{Peet}}, \bibnamefont{and}
  \bibinfo{author}{\bibfnamefont{C.}~\bibnamefont{Vafa}},
  \bibinfo{journal}{Phys. Lett.} \textbf{\bibinfo{volume}{B391}},
  \bibinfo{pages}{93} (\bibinfo{year}{1997}), \eprint{hep-th/9602065}.

\bibitem[{\citenamefont{Gauntlett et~al.}(1999)\citenamefont{Gauntlett, Myers,
  and Townsend}}]{Gauntlett:1998fz}
\bibinfo{author}{\bibfnamefont{J.~P.} \bibnamefont{Gauntlett}},
  \bibinfo{author}{\bibfnamefont{R.~C.} \bibnamefont{Myers}}, \bibnamefont{and}
  \bibinfo{author}{\bibfnamefont{P.~K.} \bibnamefont{Townsend}},
  \bibinfo{journal}{Class. Quant. Grav.} \textbf{\bibinfo{volume}{16}},
  \bibinfo{pages}{1} (\bibinfo{year}{1999}), \eprint{hep-th/9810204}.

\bibitem[{\citenamefont{Elvang et~al.}(2004)\citenamefont{Elvang, Emparan,
  Mateos, and Reall}}]{Elvang:2004rt}
\bibinfo{author}{\bibfnamefont{H.}~\bibnamefont{Elvang}},
  \bibinfo{author}{\bibfnamefont{R.}~\bibnamefont{Emparan}},
  \bibinfo{author}{\bibfnamefont{D.}~\bibnamefont{Mateos}}, \bibnamefont{and}
  \bibinfo{author}{\bibfnamefont{H.~S.} \bibnamefont{Reall}},
  \bibinfo{journal}{Phys. Rev. Lett.} \textbf{\bibinfo{volume}{93}},
  \bibinfo{pages}{211302} (\bibinfo{year}{2004}), \eprint{hep-th/0407065}.

\bibitem[{\citenamefont{Elvang et~al.}(2005{\natexlab{a}})\citenamefont{Elvang,
  Emparan, Mateos, and Reall}}]{Elvang:2004ds}
\bibinfo{author}{\bibfnamefont{H.}~\bibnamefont{Elvang}},
  \bibinfo{author}{\bibfnamefont{R.}~\bibnamefont{Emparan}},
  \bibinfo{author}{\bibfnamefont{D.}~\bibnamefont{Mateos}}, \bibnamefont{and}
  \bibinfo{author}{\bibfnamefont{H.~S.} \bibnamefont{Reall}},
  \bibinfo{journal}{Phys. Rev.} \textbf{\bibinfo{volume}{D71}},
  \bibinfo{pages}{024033} (\bibinfo{year}{2005}{\natexlab{a}}),
  \eprint{hep-th/0408120}.

\bibitem[{\citenamefont{Gauntlett and
  Gutowski}(2005{\natexlab{a}})}]{Gauntlett:2004wh}
\bibinfo{author}{\bibfnamefont{J.~P.} \bibnamefont{Gauntlett}}
  \bibnamefont{and} \bibinfo{author}{\bibfnamefont{J.~B.}
  \bibnamefont{Gutowski}}, \bibinfo{journal}{Phys. Rev.}
  \textbf{\bibinfo{volume}{D71}}, \bibinfo{pages}{025013}
  (\bibinfo{year}{2005}{\natexlab{a}}), \eprint{hep-th/0408010}.

\bibitem[{\citenamefont{Gauntlett and
  Gutowski}(2005{\natexlab{b}})}]{Gauntlett:2004qy}
\bibinfo{author}{\bibfnamefont{J.~P.} \bibnamefont{Gauntlett}}
  \bibnamefont{and} \bibinfo{author}{\bibfnamefont{J.~B.}
  \bibnamefont{Gutowski}}, \bibinfo{journal}{Phys. Rev.}
  \textbf{\bibinfo{volume}{D71}}, \bibinfo{pages}{045002}
  (\bibinfo{year}{2005}{\natexlab{b}}), \eprint{hep-th/0408122}.

\bibitem[{\citenamefont{Bena and Warner}(2004)}]{Bena:2004de}
\bibinfo{author}{\bibfnamefont{I.}~\bibnamefont{Bena}} \bibnamefont{and}
  \bibinfo{author}{\bibfnamefont{N.~P.} \bibnamefont{Warner}}
  (\bibinfo{year}{2004}), \eprint{hep-th/0408106}.

\bibitem[{\citenamefont{Dabholkar and Harvey}(1989)}]{Dabholkar:1989jt}
\bibinfo{author}{\bibfnamefont{A.}~\bibnamefont{Dabholkar}} \bibnamefont{and}
  \bibinfo{author}{\bibfnamefont{J.~A.} \bibnamefont{Harvey}},
  \bibinfo{journal}{Phys. Rev. Lett.} \textbf{\bibinfo{volume}{63}},
  \bibinfo{pages}{478} (\bibinfo{year}{1989}).

\bibitem[{\citenamefont{Dabholkar et~al.}(1990)\citenamefont{Dabholkar,
  Gibbons, Harvey, and Ruiz~Ruiz}}]{Dabholkar:1990yf}
\bibinfo{author}{\bibfnamefont{A.}~\bibnamefont{Dabholkar}},
  \bibinfo{author}{\bibfnamefont{G.~W.} \bibnamefont{Gibbons}},
  \bibinfo{author}{\bibfnamefont{J.~A.} \bibnamefont{Harvey}},
  \bibnamefont{and}
  \bibinfo{author}{\bibfnamefont{F.}~\bibnamefont{Ruiz~Ruiz}},
  \bibinfo{journal}{Nucl. Phys.} \textbf{\bibinfo{volume}{B340}},
  \bibinfo{pages}{33} (\bibinfo{year}{1990}).

\bibitem[{\citenamefont{Gaiotto
  et~al.}(2005{\natexlab{a}})\citenamefont{Gaiotto, Strominger, and
  Yin}}]{Gaiotto:2005gf}
\bibinfo{author}{\bibfnamefont{D.}~\bibnamefont{Gaiotto}},
  \bibinfo{author}{\bibfnamefont{A.}~\bibnamefont{Strominger}},
  \bibnamefont{and} \bibinfo{author}{\bibfnamefont{X.}~\bibnamefont{Yin}}
  (\bibinfo{year}{2005}{\natexlab{a}}), \eprint{hep-th/0503217}.

\bibitem[{\citenamefont{Elvang et~al.}(2005{\natexlab{b}})\citenamefont{Elvang,
  Emparan, Mateos, and Reall}}]{Elvang:2005sa}
\bibinfo{author}{\bibfnamefont{H.}~\bibnamefont{Elvang}},
  \bibinfo{author}{\bibfnamefont{R.}~\bibnamefont{Emparan}},
  \bibinfo{author}{\bibfnamefont{D.}~\bibnamefont{Mateos}}, \bibnamefont{and}
  \bibinfo{author}{\bibfnamefont{H.~S.} \bibnamefont{Reall}},
  \bibinfo{journal}{JHEP} \textbf{\bibinfo{volume}{08}}, \bibinfo{pages}{042}
  (\bibinfo{year}{2005}{\natexlab{b}}), \eprint{hep-th/0504125}.

\bibitem[{\citenamefont{Gaiotto
  et~al.}(2005{\natexlab{b}})\citenamefont{Gaiotto, Strominger, and
  Yin}}]{Gaiotto:2005xt}
\bibinfo{author}{\bibfnamefont{D.}~\bibnamefont{Gaiotto}},
  \bibinfo{author}{\bibfnamefont{A.}~\bibnamefont{Strominger}},
  \bibnamefont{and} \bibinfo{author}{\bibfnamefont{X.}~\bibnamefont{Yin}}
  (\bibinfo{year}{2005}{\natexlab{b}}), \eprint{hep-th/0504126}.

\bibitem[{\citenamefont{Bena et~al.}(2005)\citenamefont{Bena, Kraus, and
  Warner}}]{Bena:2005ni}
\bibinfo{author}{\bibfnamefont{I.}~\bibnamefont{Bena}},
  \bibinfo{author}{\bibfnamefont{P.}~\bibnamefont{Kraus}}, \bibnamefont{and}
  \bibinfo{author}{\bibfnamefont{N.~P.} \bibnamefont{Warner}}
  (\bibinfo{year}{2005}), \eprint{hep-th/0504142}.

\bibitem[{\citenamefont{Iizuka and Shigemori}(2005)}]{Iizuka:2005uv}
\bibinfo{author}{\bibfnamefont{N.}~\bibnamefont{Iizuka}} \bibnamefont{and}
  \bibinfo{author}{\bibfnamefont{M.}~\bibnamefont{Shigemori}},
  \bibinfo{journal}{JHEP} \textbf{\bibinfo{volume}{08}}, \bibinfo{pages}{100}
  (\bibinfo{year}{2005}), \eprint{hep-th/0506215}.

\bibitem[{\citenamefont{Behrndt et~al.}(2005)\citenamefont{Behrndt, Cardoso,
  and Mahapatra}}]{Behrndt:2005he}
\bibinfo{author}{\bibfnamefont{K.}~\bibnamefont{Behrndt}},
  \bibinfo{author}{\bibfnamefont{G.~L.} \bibnamefont{Cardoso}},
  \bibnamefont{and} \bibinfo{author}{\bibfnamefont{S.}~\bibnamefont{Mahapatra}}
  (\bibinfo{year}{2005}), \eprint{hep-th/0506251}.

\bibitem[{\citenamefont{Dabholkar et~al.}(1996)\citenamefont{Dabholkar,
  Gauntlett, Harvey, and Waldram}}]{Dabholkar:1995nc}
\bibinfo{author}{\bibfnamefont{A.}~\bibnamefont{Dabholkar}},
  \bibinfo{author}{\bibfnamefont{J.~P.} \bibnamefont{Gauntlett}},
  \bibinfo{author}{\bibfnamefont{J.~A.} \bibnamefont{Harvey}},
  \bibnamefont{and} \bibinfo{author}{\bibfnamefont{D.}~\bibnamefont{Waldram}},
  \bibinfo{journal}{Nucl. Phys.} \textbf{\bibinfo{volume}{B474}},
  \bibinfo{pages}{85} (\bibinfo{year}{1996}), \eprint{hep-th/9511053}.

\bibitem[{\citenamefont{Callan et~al.}(1996)\citenamefont{Callan, Maldacena,
  and Peet}}]{Callan:1995hn}
\bibinfo{author}{\bibfnamefont{C.~G.} \bibnamefont{Callan}},
  \bibinfo{author}{\bibfnamefont{J.~M.} \bibnamefont{Maldacena}},
  \bibnamefont{and} \bibinfo{author}{\bibfnamefont{A.~W.} \bibnamefont{Peet}},
  \bibinfo{journal}{Nucl. Phys.} \textbf{\bibinfo{volume}{B475}},
  \bibinfo{pages}{645} (\bibinfo{year}{1996}), \eprint{hep-th/9510134}.

\bibitem[{\citenamefont{Lunin and Mathur}(2001)}]{Lunin:2001fv}
\bibinfo{author}{\bibfnamefont{O.}~\bibnamefont{Lunin}} \bibnamefont{and}
  \bibinfo{author}{\bibfnamefont{S.~D.} \bibnamefont{Mathur}},
  \bibinfo{journal}{Nucl. Phys.} \textbf{\bibinfo{volume}{B610}},
  \bibinfo{pages}{49} (\bibinfo{year}{2001}), \eprint{hep-th/0105136}.

\bibitem[{\citenamefont{Lunin and Mathur}(2002)}]{Lunin:2002qf}
\bibinfo{author}{\bibfnamefont{O.}~\bibnamefont{Lunin}} \bibnamefont{and}
  \bibinfo{author}{\bibfnamefont{S.~D.} \bibnamefont{Mathur}},
  \bibinfo{journal}{Phys. Rev. Lett.} \textbf{\bibinfo{volume}{88}},
  \bibinfo{pages}{211303} (\bibinfo{year}{2002}), \eprint{hep-th/0202072}.

\bibitem[{\citenamefont{Denef}(2000)}]{Denef:2000nb}
\bibinfo{author}{\bibfnamefont{F.}~\bibnamefont{Denef}},
  \bibinfo{journal}{JHEP} \textbf{\bibinfo{volume}{08}}, \bibinfo{pages}{050}
  (\bibinfo{year}{2000}), \eprint{hep-th/0005049}.

\bibitem[{\citenamefont{Palmer and Marolf}(2004)}]{Palmer:2004gu}
\bibinfo{author}{\bibfnamefont{B.~C.} \bibnamefont{Palmer}} \bibnamefont{and}
  \bibinfo{author}{\bibfnamefont{D.}~\bibnamefont{Marolf}},
  \bibinfo{journal}{JHEP} \textbf{\bibinfo{volume}{06}}, \bibinfo{pages}{028}
  (\bibinfo{year}{2004}), \eprint{hep-th/0403025}.

\bibitem[{\citenamefont{Marolf and Palmer}(2004)}]{Marolf:2004fy}
\bibinfo{author}{\bibfnamefont{D.}~\bibnamefont{Marolf}} \bibnamefont{and}
  \bibinfo{author}{\bibfnamefont{B.~C.} \bibnamefont{Palmer}},
  \bibinfo{journal}{Phys. Rev.} \textbf{\bibinfo{volume}{D70}},
  \bibinfo{pages}{084045} (\bibinfo{year}{2004}), \eprint{hep-th/0404139}.

\bibitem[{\citenamefont{Bena and Kraus}(2004)}]{Bena:2004tk}
\bibinfo{author}{\bibfnamefont{I.}~\bibnamefont{Bena}} \bibnamefont{and}
  \bibinfo{author}{\bibfnamefont{P.}~\bibnamefont{Kraus}},
  \bibinfo{journal}{JHEP} \textbf{\bibinfo{volume}{12}}, \bibinfo{pages}{070}
  (\bibinfo{year}{2004}), \eprint{hep-th/0408186}.

\bibitem[{\citenamefont{Kraus and Larsen}(2005)}]{Kraus:2005vz}
\bibinfo{author}{\bibfnamefont{P.}~\bibnamefont{Kraus}} \bibnamefont{and}
  \bibinfo{author}{\bibfnamefont{F.}~\bibnamefont{Larsen}},
  \bibinfo{journal}{JHEP} \textbf{\bibinfo{volume}{09}}, \bibinfo{pages}{034}
  (\bibinfo{year}{2005}), \eprint{hep-th/0506176}.

\bibitem[{\citenamefont{Bak et~al.}(2005)\citenamefont{Bak, Hyakutake, Kim, and
  Ohta}}]{Bak:2004kz}
\bibinfo{author}{\bibfnamefont{D.}~\bibnamefont{Bak}},
  \bibinfo{author}{\bibfnamefont{Y.}~\bibnamefont{Hyakutake}},
  \bibinfo{author}{\bibfnamefont{S.}~\bibnamefont{Kim}}, \bibnamefont{and}
  \bibinfo{author}{\bibfnamefont{N.}~\bibnamefont{Ohta}},
  \bibinfo{journal}{Nucl. Phys.} \textbf{\bibinfo{volume}{B712}},
  \bibinfo{pages}{115} (\bibinfo{year}{2005}), \eprint{hep-th/0407253}.

\bibitem[{\citenamefont{Russo and Susskind}(1995)}]{Russo:1994ev}
\bibinfo{author}{\bibfnamefont{J.~G.} \bibnamefont{Russo}} \bibnamefont{and}
  \bibinfo{author}{\bibfnamefont{L.}~\bibnamefont{Susskind}},
  \bibinfo{journal}{Nucl. Phys.} \textbf{\bibinfo{volume}{B437}},
  \bibinfo{pages}{611} (\bibinfo{year}{1995}), \eprint{hep-th/9405117}.

\bibitem[{\citenamefont{Dabholkar et~al.}(2005, to
  appear)\citenamefont{Dabholkar, Iizuka, Iqubal, Sen, and
  Shigemori}}]{Progress}
\bibinfo{author}{\bibfnamefont{A.}~\bibnamefont{Dabholkar}},
  \bibinfo{author}{\bibfnamefont{N.}~\bibnamefont{Iizuka}},
  \bibinfo{author}{\bibfnamefont{A.}~\bibnamefont{Iqubal}},
  \bibinfo{author}{\bibfnamefont{A.}~\bibnamefont{Sen}}, \bibnamefont{and}
  \bibinfo{author}{\bibfnamefont{M.}~\bibnamefont{Shigemori}}
  (\bibinfo{year}{2005, to appear}).

\end{thebibliography}
\bibliographystyle{apsrev}

\end{document}